\documentclass[acus]{JAC2003}
\usepackage{graphicx}
\usepackage{booktabs}
\usepackage[center]{subfigure}

\begin{document}
\title{STATUS OF THE HIE-ISOLDE PROJECT AT CERN}
\author{M.A. Fraser, Y. Kadi, A.P. Bernardes, Y. Blumenfeld, E. Bravin, S. Calatroni, R. Catherall, \\ B. Goddard, D. Parchet, E. Siesling, W. Venturini Delsolaro, G. Vandoni, D. Voulot, L.R. Williams \\ \\ CERN, Geneva, Switzerland}

\maketitle

\begin{abstract}

The HIE-ISOLDE project represents a major upgrade of the ISOLDE nuclear facility with a mandate to significantly improve the quality and increase the intensity and energy of radioactive nuclear beams produced at CERN. The project will expand the experimental nuclear physics programme at ISOLDE by focusing on an upgrade of the existing Radioactive ion beam EXperiment (REX) linac with a 40~MV superconducting linac comprising thirty-two niobium-on-copper sputter-coated quarter-wave resonators housed in six cryomodules. The new linac will raise the energy of post-accelerated beams from 3~MeV/$u$ to over 10~MeV/$u$. The upgrade will be staged to first deliver beam energies of 5.5~MeV/$u$ using two high-$\beta$ cryomodules placed downstream of REX, before the energy variable section of the existing linac is replaced with two low-$\beta$ cryomodules and two additional high-$\beta$ cryomodules are installed to attain over 10~MeV/$u$ with full energy variability above 0.45~MeV/$u$. An overview of the project including a status summary of the different R\&D activities and the schedule will outlined.

\end{abstract}

\section{INTRODUCTION}

The High Intensity and Energy (HIE) ISOLDE project~\cite{HIEISOLDE} aims at several important upgrades of the present ISOLDE radioactive beam facility at CERN. The main focus lies in the energy upgrade of the post-accelerated radionuclide beams from 3 MeV/$u$ up to over 10 MeV/$u$ through the addition of superconducting (SC) quarter-wave resonators (QWRs) operating at 101.28 MHz. This will open the possibility of many new types of experiments including transfer reactions throughout the nuclear chart.

The project also includes a design study that aims at improving the target and front-end part of ISOLDE to fully benefit from upgrades of the existing CERN proton injectors, e.g. LINAC4 and upgrade in energy of the PS Booster. This improvement combined with upgrades to the RILIS laser ion source and the radiofrequency quadrupole (RFQ) cooler and buncher (ISCOOL) will lead to an increase of radioactive beam intensities of up to an order of magnitude. The beam emittance will be improved with the implementation of ISCOOL placed after a pre-separator but before a new High-Resolution Separator (HRS). The new HRS, based on the latest magnet technology, will have sufficient mass resolution to permit isobaric separation. ISCOOL will also permit a tailoring of the time structure of the beam, removing the dependence on the proton beam time structure and diffusion-effusion properties of the target and ion source units. Highly charged ions will be provided for REX and other users through an improved low energy stage of REX-ISOLDE and a possible installation of an upgraded Electron Beam Ion Source (EBIS) charge breeder.

\begin{figure}[ht!]
   \centering
   \includegraphics[width=84mm]{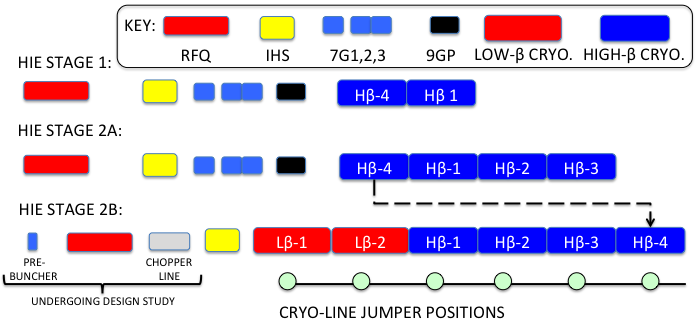}
   \caption{A schematic of the staged installation of the HIE-ISOLDE linac (Existing REX structures: RFQ, IHS: 20-gap IH-structure, 7GX: 7-gap split-ring cavities, 9GP: 9-gap IH-structure).}
   \label{stage_schematic}
\end{figure}

The linac upgrade will be staged in order to deliver higher beam energies to the experiments as soon as possible, with future upgrade stages ensuring a wide range of energy variability and providing an optional $\sim$100~ns bunch spacing. The first stage of the upgrade involves the design, construction, installation and commissioning of two cryomodules downstream of REX, the existing post-accelerator. These cryomodules will each house five \mbox{high-$\beta$} ($\beta_g = 10.3\%$) SC cavities and one SC solenoid. Extra cryomodules will be added to the beam line in a modular fashion until all six cryomodules, including two cryomodules housing six low-$\beta$ ($\beta_g = 6.3\%$) SC cavities and two SC solenoids, are online. The upgrade will be completed with a final stage that will see the linac extended in order to make room to pre-bunch the beam into the existing RFQ accelerator at a sub-harmonic frequency below 101.28~MHz, allowing the bunch spacing to be increased without significant loss in transmission; time-of-flight particle detection will then be viable at the experiments. Also foreseen is a beam chopper to reject the background of populated satellite bunches either side of the main sub-harmonic beam pulses. The staged installation of the linac is shown schematically in Figure~\ref{stage_schematic}.
The pre-bunching/chopping scheme is under study and still to be finalised, e.g. chopping could occur before RFQ instead of after.

The beam dynamics studies \cite{physrev_fraser, fraser_phd} arrived at a compact lattice, based on cavities operating at a gradient of 6~MV/m,\footnote{Cavity active length definitions: $\textrm{low-}\beta = 195~\textrm{mm}$ and $\textrm{high-}\beta = 300~\textrm{mm}$.} designed both due to the limited space in the experimental hall and to maximise the dynamical acceptance and optical performance of the machine. These considerations influenced the specification of the SC solenoids, alignment system, beam diagnostics system, steering magnets and cryomodules. 

\begin{figure}[h]
\centering
\mbox{
\hspace{-0.65cm}
\subfigure[Stages 1 and 2a]{
\includegraphics[scale=0.40]{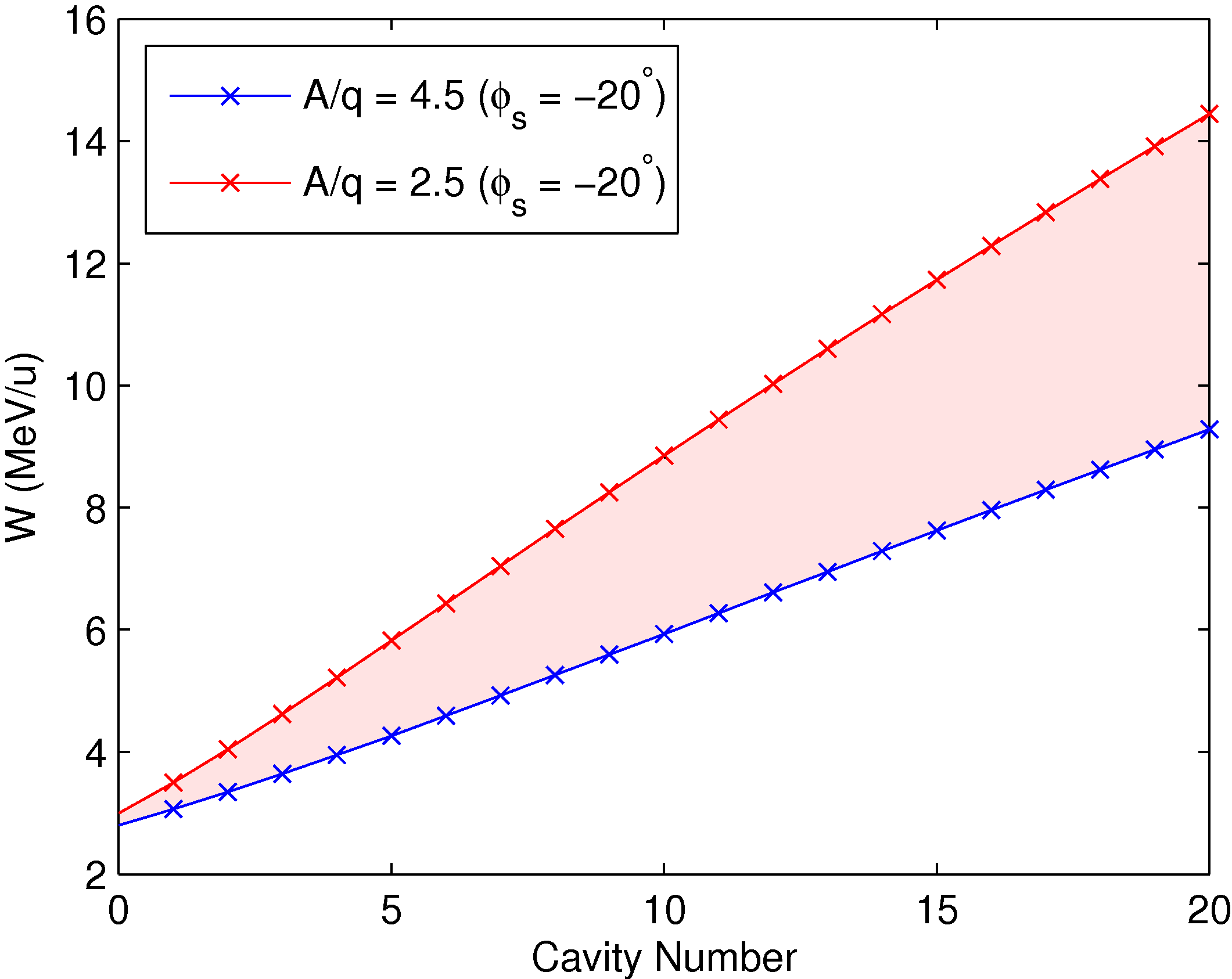}
}
\subfigure[Stage 2b]{
\includegraphics[scale=0.40]{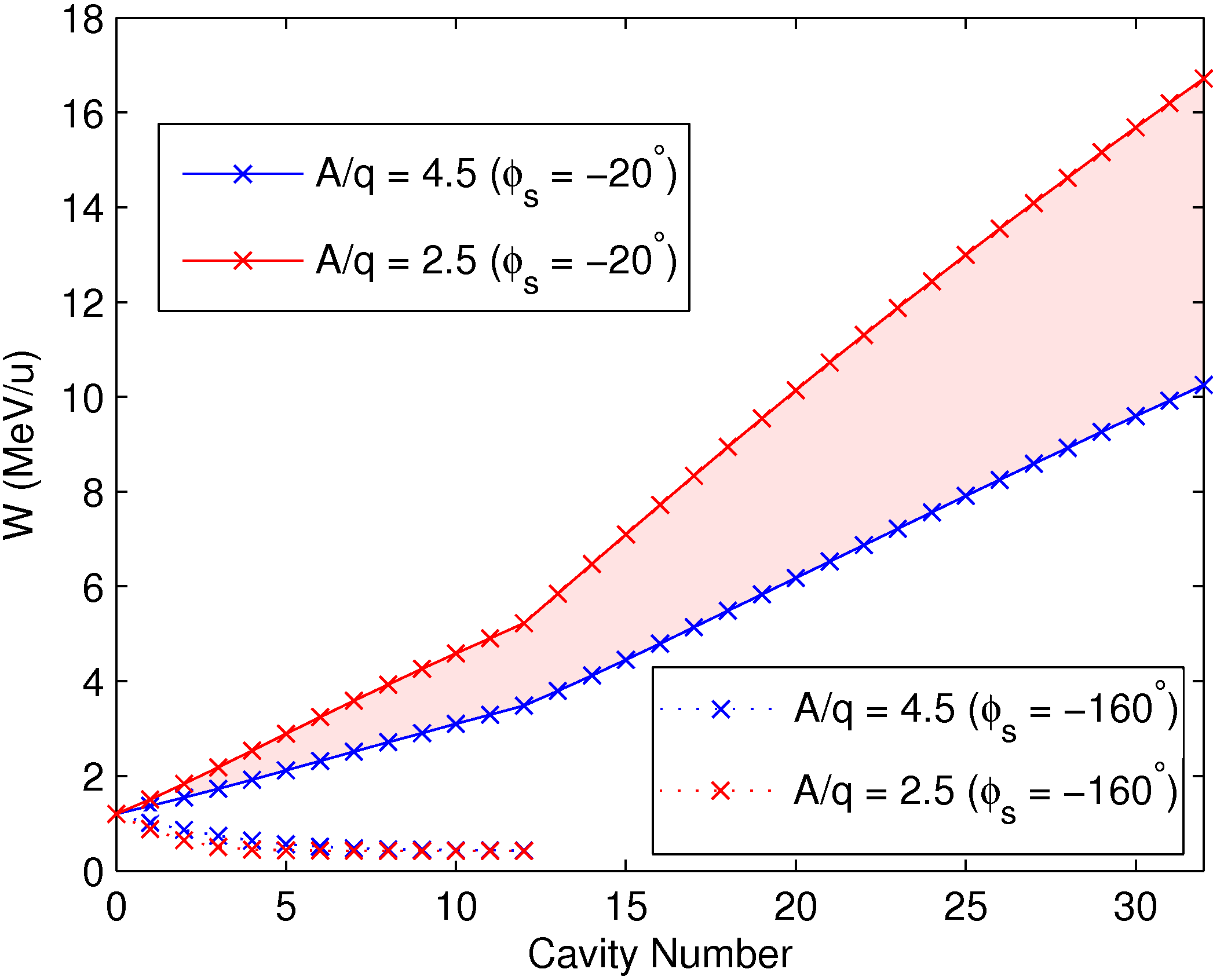}
}}
\caption{Energy reach of post-accelerated beams for each stage of the upgrade as a function of the $A/q$ acceptance.}
\label{energy_stages}
\end{figure}

The superconducting linac is designed to provide a total accelerating voltage of 39.6~MV with an average synchronous phase $\phi_s$ of $-20$~deg, i.e. the minimum voltage required in order to achieve a final energy of at least 10~MeV/$u$ with the heaviest beams that have a mass-to-charge state $A/q = 4.5$. The normal conducting 20-gap IH-structure (IHS) limits the $A/q$ acceptance of the machine to $2.5 \leq A/q \leq 4.5$. The energy range accessible with the new SC linac is shown for the each stage in Figure~\ref{energy_stages}, including the decelerated beam that opens a range of energy not previously accessible with REX because of the fixed velocity profile of the IHS.

A research and development programme looking at all the different aspects of the SC linac started in 2008 and continued throughout 2011. In particular, the R\&D effort has focused on the development of the high-$\beta$ cavity, for which it has been decided to adopt technology based on copper cavities sputter-coated with niobium. The required infrastructure has been determined and the integration of the SC linac and High Energy Beam Transfer (HEBT) lines inside the existing experimental hall has been finalised.

The project has been approved by CERN and its implementation started in January 2010. An overview of the project and its timeline will be given below.

\section{PHYSICS MOTIVATION}

The current REX maximum energy of 3~MeV/$u$ largely restricts the physics program to measurements of collective degrees of freedom through single-step Coulomb excitation and probing single-particle degrees of freedom of light nuclei through direct nuclear reactions. The provision of exotic beams with energies up to 10~MeV/$u$ will greatly expand the application of these methods to all ISOLDE beams~\cite{HIE_sci}. Complete measurements of low-lying collective structure will be possible and, for particle states, unambiguous physical quantities can be extracted. The versatility of the accelerator, with variable energy down to 0.45~MeV/$u$, will facilitate measurements of rates of reactions that are the building blocks of nucleosynthesis chains, while the higher energy beams from the new linac will also allow prototype studies of new collective modes using electromagnetic probes and studies of exotic proton-rich and neutron-rich nuclei using fusion and highly-damped binary collisions.

The installation of the TSR storage ring~\cite{tsr} at HIE-ISOLDE has been approved and provides unique opportunities by being located at an ISOL facility. Reaction experiments could benefit from an increased luminosity from being in-ring, through multiple beam passes ($\sim$1~MHz), or from extracting electron-cooled beams for ultra-high resolution studies of heavy nuclei and Coulex studies. The TSR could also be used as a beam stretcher to accumulate the pulsed beam delivered by the EBIS and extract it as a d.c. beam over a few seconds, alleviating problems from the high instantaneous rates encountered with pulsed beams.

The 34 letters of intent submitted in 2010 for experiments at HIE-ISOLDE include 284 participants. The demand for the linac upgrade is reflected in 88\% of the letters requesting higher beam energies that will be provided by the HIE upgrade.

\section{INTEGRATION AT ISOLDE}

\begin{figure}[ht]
   \centering
   \includegraphics[width=82mm]{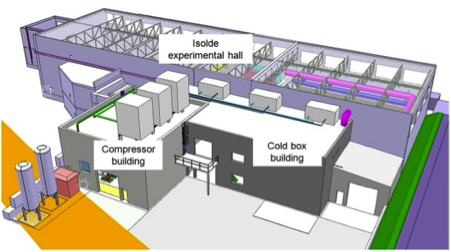}
   \caption{New HIE-ISOLDE buildings.}
   \label{buildings}
\end{figure}

The new HIE-ISOLDE SC linac will require a major increase of equipment to the existing facility's infrastructure. Two new surface buildings will be constructed in order to house the helium compressor station and the helium refrigerator cold box, see Figure~\ref{buildings}.

\begin{figure}[ht]
   \centering
   \includegraphics[width=82mm]{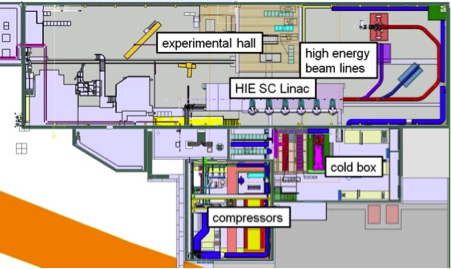}
   \caption{Layout of the HIE-ISOLDE facility.}
   \label{hie_layout}
\end{figure}

Ground breaking started at the end of last summer (2011) with the preparation of the site and the construction of a new secure access point to the experimental hall for personnel and material while HIE-ISOLDE construction work is on-going. Presently, civil engineering works are in progress for the construction of the new buildings adjacent to the existing ISOLDE experimental hall. The construction works should be completed by the third quarter of 2012 after which the installation of the electrical systems and main services will take over.
The cryogenic station installation, shown in Figure~\ref{hie_layout}, will start in the second quarter of 2013. The He liquefier will be installed in a separate light construction building as close as possible to the linac in order to minimise the length of the LHe distribution line. This will enable an easier and more stable operation of the cryogenic  system. The cryogenic system includes a cryogenic transfer line that will link the cold box to the different interconnecting ``jumper'' boxes, feeding from the top, the six cryomodules of the new SC linac.

\begin{figure}[ht]
   \centering
   \includegraphics[width=82mm]{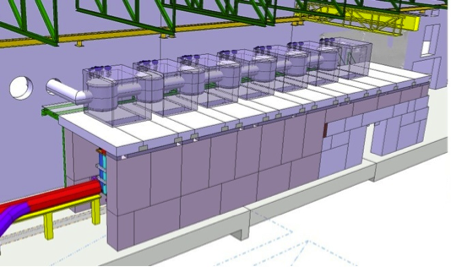}
   \caption{Linac shielding tunnel and cryogenic transfer line.}
   \label{tunnel}
\end{figure}

The shielding tunnel will be installed in its entirety for the full linac along with the cryogenic transfer line, see Figure~\ref{tunnel}, whereas the linac and high energy beam transfer lines will be installed in stages. A new HEBT, discussed below, will bring the beam into the existing extension of the ISOLDE experimental hall~\cite{ipac12_op}.

\section{CRYOMODULE DESIGN}

\begin{figure}[h]
   \centering
   \includegraphics[width=85mm]{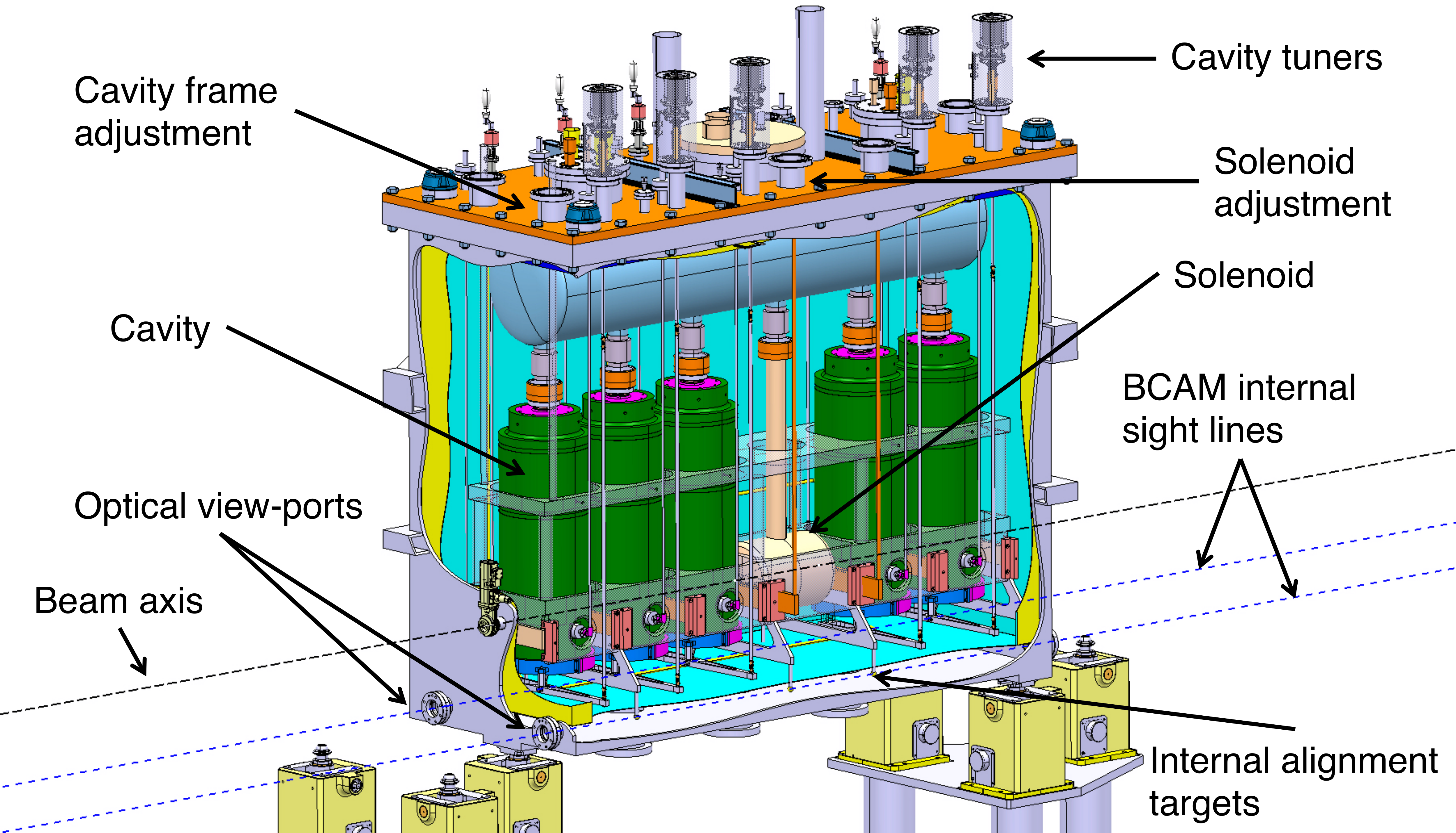}
   \caption{3D model of the high-$\beta$ cryomodule (beam direction is from right to left).}
   \label{cryo_pic}
\end{figure}

The cavities will be grouped in common cryomodules with six cavities and two focusing solenoids for the low-$\beta$ cryomodules, and five cavities and one focusing solenoid for the high-$\beta$ cryomodules. Figure~\ref{cryo_pic} shows a 3D model of the high-$\beta$ cryomodule. In order to simplify the mechanical design and assembly, and to minimise the drift length between cavities and the overall length of the machine, a common vacuum was chosen for the beam and cryogenic insulation. The solenoids need to be aligned with a stringent precision of $\pm0.15$~mm (i.e.~$\pm1\sigma$, where the simulated Gaussian error distribution was truncated at $\pm3\sigma$) and a system of independent adjustment, useable under vacuum and at operational temperatures, of the solenoid with respect to the cavity string is foreseen. An active position monitoring system based on BCAM CCD cameras~\cite{brandeis} is under development. The active components will be cooled to 4.5~K in two stages using gaseous and liquid He. Insulation will be guaranteed by a heat screen at 75~K. A vacuum of $10^{-8}$~mbar after cryopumping is necessary for optimal operation.
To ensure the cleanest vacuum conditions, the vacuum system will be entirely dry, with turbo-molecular pumps backed by dry scrolls. In addition, automatic procedures with staged limitation of conductance, both for slow pump-down in the viscous regime and for venting with dry nitrogen, will be applied. A detailed description of the high-$\beta$ cryomodule design can be found in~\cite{ipac11_cryo}.

\section{CAVITY DEVELOPMENT}

The development of the SC cavities started in 2008 and focused on the high-$\beta$ cavity design, which will be installed first. The cavities are based on the niobium sputtered copper technology pioneered at INFN-LNL. They are specified to reach a nominal accelerating field of 6~MV/m on axis over an active length of 300~mm. A $Q$-value of $5\times 10^8$ is necessary to reach this field with a power dissipation of 10~W. In recent years much effort has been put in the design of a coating facility and the optimisation of the coating process. Since 2011 a single-cavity test-stand has been operational at CERN \cite{ipac11_test_stand} working in tandem with the sputtering development to improve the quality factor of the niobium sputtered cavities. The latest tests have shown $Q$-values of $1.5\times10^8$ at 6~MV/m; close to the design goal. Further details can be found in~\cite{ipac11_cavity}.

\section{BEAM DIAGNOSTICS}

\begin{figure}[h!]
   \centering
   \includegraphics[width=80mm]{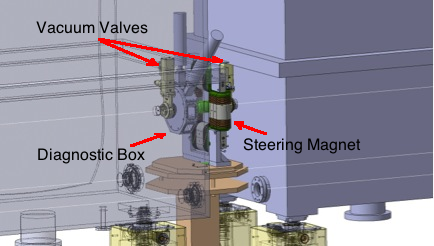}
   \caption{3D model of the region between cryomodules.}
   \label{inter_tank}
\end{figure}

A beam instrumentation R\&D programme is ongoing to provide a solution for the HIE-ISOLDE beam diagnostic system~\cite{ipac12_bd}. The space available in the regions between the cryomodules, in which a dipole steering magnet and vacuum valves are also situated, is very constrained, see Figure~\ref{inter_tank}. Dimensioned at 58 mm in the longitudinal direction, the box is extremely compact and is designed to operate in the stray-field of the adjacent steering magnet. More details of the air-cooled steering magnet design specified at 6~T~mm can be found in~\cite{bauche}.

\begin{figure}[ht]
   \centering
   \includegraphics[width=80mm]{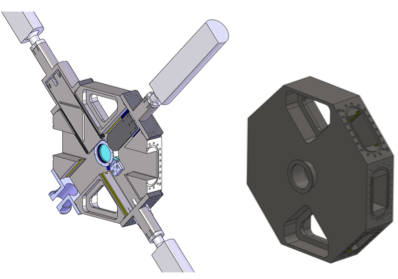}
   \caption{3D model of the short diagnostic box, courtesy of AVS technology.}
   \label{avs_box}
\end{figure}

An octagonal solution for the main body of the diagnostic box was chosen to facilitate multiple functionalities in the compact space available for the box, see Figure~\ref{avs_box}. The same design will be used in the HEBT lines, numbering fifteen units in the first stage of the upgrade. Up to five different devices can be actuated in-and-out of the beam in the transverse plane, with a beam profiling functionality included in two of these five devices as standard: a scanning slit that profiles the beam in front of a Faraday cup. The Faraday cup has been designed with a length of just 20~mm and a sensitivity from $1 - 500$~pA. The prototype system will be delivered to CERN by Added Value Solutions (AVS) this summer for testing of the Faraday cup with beam before the Long Shutdown of the CERN accelerator complex commences (December 2012 until April 2014).

A longitudinal beam diagnostics system based on a silicon detector has been developed for the fast, and eventually automated, phasing of the large number of cavities that will accompany the upgrade. The system is capable of providing both relative phase and energy measurements. The proof-of-principle has been tested at REX by phasing the third 7-gap resonator with the system. For more details the reader is referred to~\cite{zocca}. A time-of-flight system for absolute energy measurements is also being considered. 

\section{COMMISSIONING AND OPERATION}

\begin{figure}[h]
\centering
\mbox{
\hspace{-0.5cm}
\subfigure[Stage 1 ($W_0 = 2.85~\textrm{MeV}/u$).]{
\includegraphics[scale=0.44]{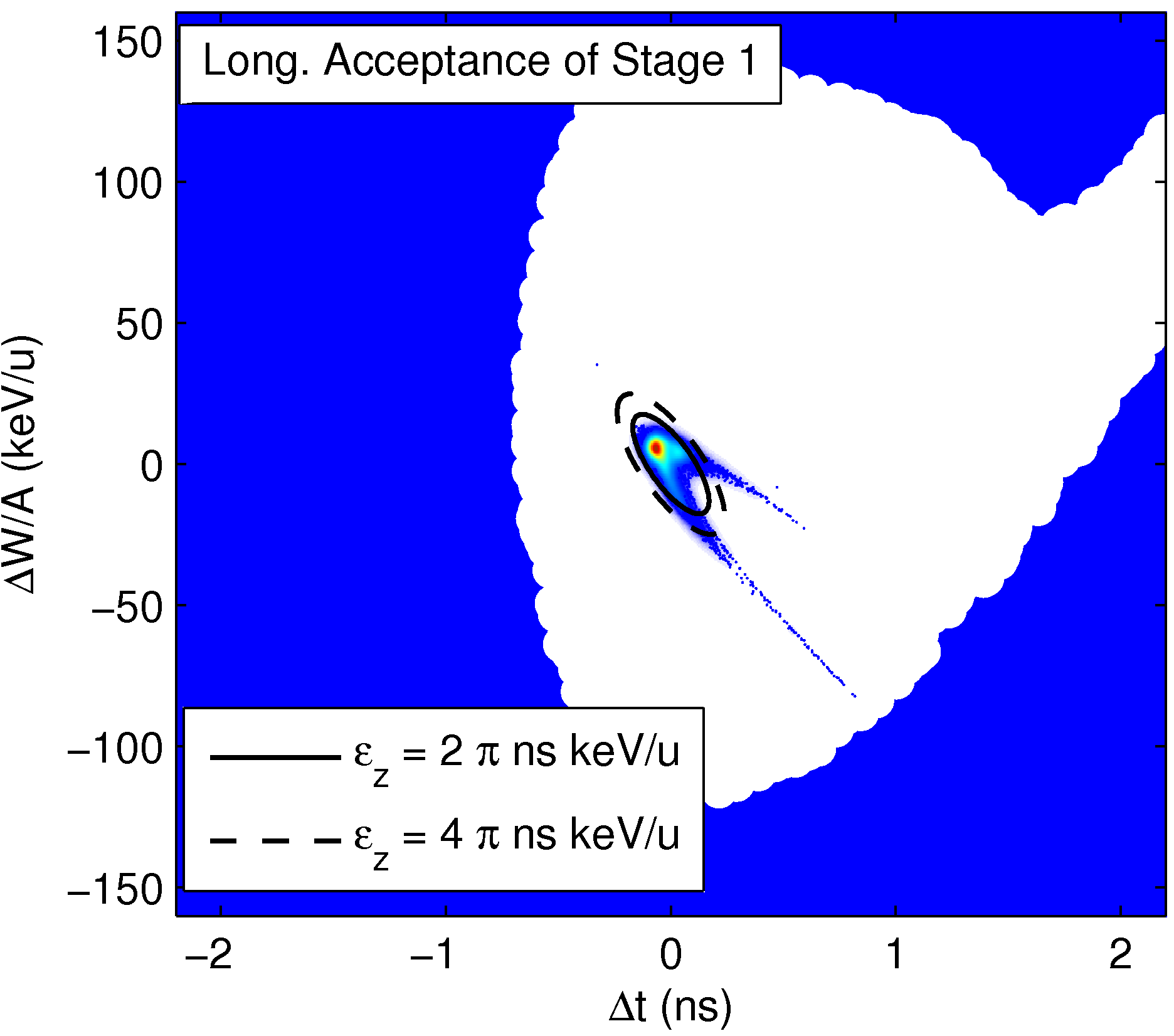}
}
\subfigure[Stage 2b ($W_0 = 1.2~\textrm{MeV}/u$).]{
\includegraphics[scale=0.44]{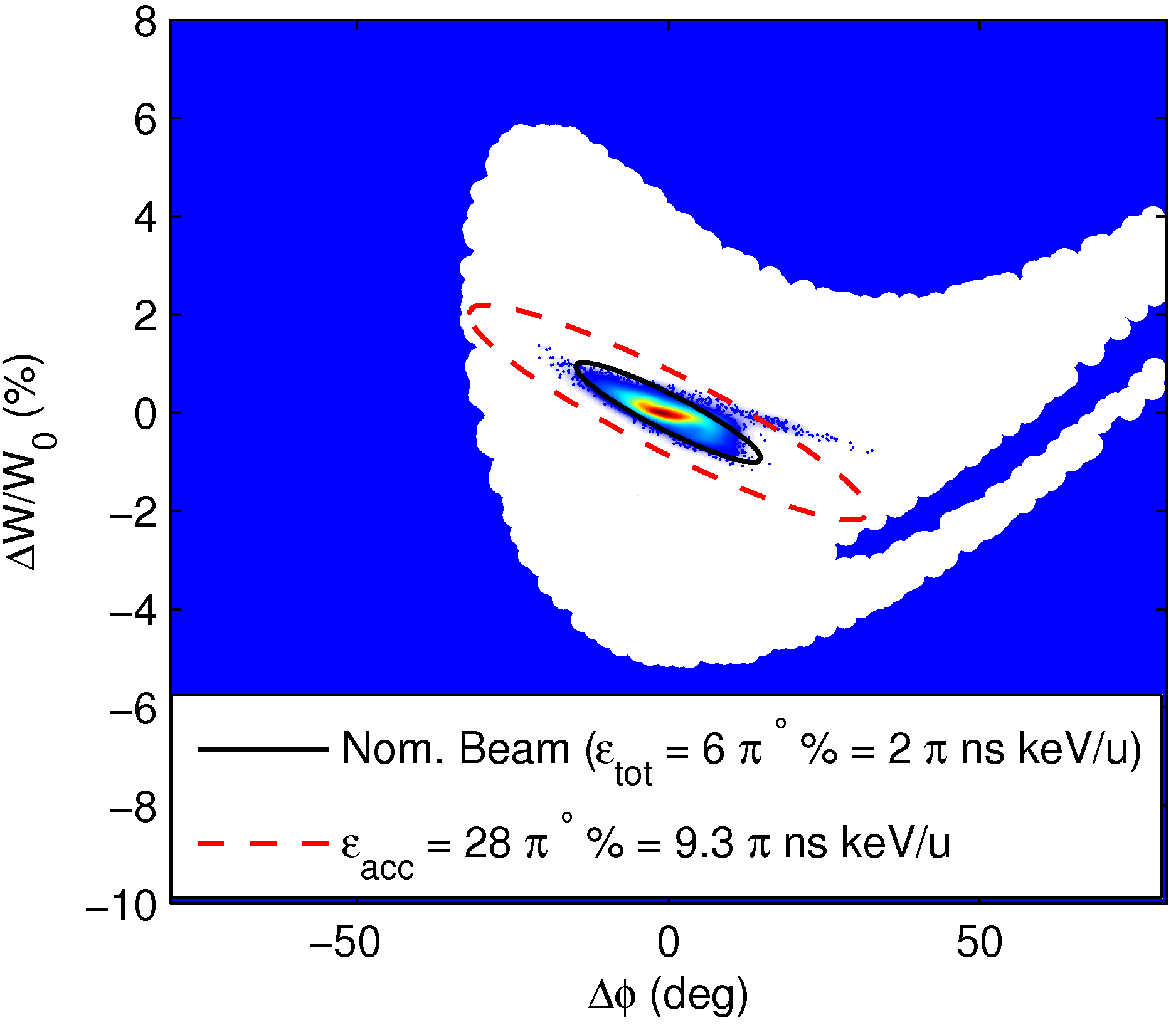}
}
}
\caption{Simulated beam distribution (consistent with measurements) in longitudinal phase-space at entry to the SC linac, compared to the acceptance (shown in white).}
\label{acceptance}
\end{figure}

In preparation for the linac upgrade various recommissioning steps have been undertaken to better understand the optimum working points of the existing accelerating structures, including bead-pull measurements, emittance measurements and end-to-end beam dynamics simulations benchmarked to measurements, described in~\cite{fraser_phd}. The emittance delivered by REX was shown to be compatible with the acceptance of the superconducting linac in both the transverse and longitudinal phase-space planes, see~\cite{voulot_emit} and~\cite{fraser_emit} for the details of the respective measurements. The longitudinal acceptance of the SC linac is compared to the phase-space of the beam delivered by REX in Stages 1 and 2b of the upgrade in Figure~\ref{acceptance}. The normalised 90\% transverse emittance was measured as less than 0.3~$\pi\textrm{ mm mrad}$ and the 86\% longitudinal emittance was measured as 1.5~$\pi\textrm{ ns keV/}u$.

To date, transverse emittance measurements using the NTG emittance rig~\cite{ntg} have been limited to those made on high intensity ($\sim\textrm{nA}$) pilot beams, which are generated by leaking gas into the EBIS. It is likely that the beam emittance is degraded with respect to the radioactive beam emittance due to the disruptive space-charge effects produced by the higher beam intensity and resulting compensation of the radial focusing force provided to the trapped ions in the electron beam of the source. For this reason a high sensitivity emittance measurement system using single-particle detectors is being developed to measure the radioactive beam emittance using slits and detectors housed in adjacent diagnostic boxes that were shown previously. An emittance measurement will also be attempted this summer at lower beam intensity ($\sim$10~pA) by avoiding the use of slits to sample the phase-space in the emittance rig and instead using the three-gradient method to reconstruct the phase-space distribution by tomography; the beam size at the wire-grid of the emittance rig will be varied as a function of the gradient of an upstream quadrupole.

Due to the short nature of REX experimental runs (typically three to ten days) and the need for frequent energy and beam changes dictated by the physics programme, REX operation requires fast and reliable set-ups and the possibility to switch between beams or change the energy within a few hours. The size and complexity of the machine will increase with the energy upgrade, e.g. the number of cavities will increase from seven in the present linac to thirty-five in the final version of the HIE linac. The cavity phases will also become dependent on $A/q$ as the full voltage provided by the QWRs can be exploited and the velocity profile changes as a function of $A/q$ in the SC part of the linac. For this reason an automatic phasing procedure is foreseen using the above mentioned silicon detector either by tracking the relative energy change or, when the 100~ns bunch spacing becomes available, using a time-of-flight method.

Software applications for the control system are being developed in preparation for the commissioning phase, including the possibility of setting machine parameter values directly from beam simulation codes and vice versa. Automated optimisation software is also being investigated to phase the linac quickly and some initial tests have already been done on the existing linac. 

\section{HIGH ENERGY BEAM TRANSFER LINES AND EXPERIMENTAL AREAS}

The Miniball segmented germanium array and T-REX experimental setups already operational at REX will be used intensively when the first beams are delivered in 2015.
\begin{figure}[h!]
   \centering
   \includegraphics[width=80mm]{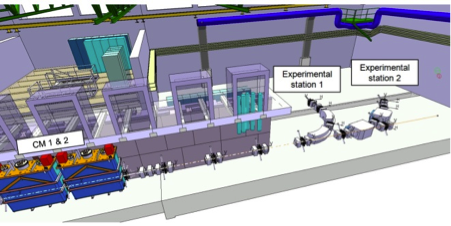}
   \caption{Stage 1 of the HIE upgrade with two high-$\beta$ cryomodules and HEBT lines added to REX.}
   \label{hie_stage1_hall}
\end{figure}
Only the first two high-$\beta$ cryomodules will be installed in the tunnel, downstream of the existing REX linac and an open beam line will also be available for traveling experiments, see Figure~\ref{hie_stage1_hall}.

The following year it is foreseen to install a helical-orbit spectrometer that will focus on transfer reactions across the entire nuclear chart, made possible by the upgraded beam energy. The helical-orbit spectrometer will demand a bunch spacing of the order of 100~ns and a scheme to pre-bunch before the RFQ at a sub-harmonic of its natural frequency is foreseen to satisfy this requirement. An active-target experiment for resonant scattering and transfer reactions is also under consideration. 

\begin{figure}[h!]
   \centering
   \includegraphics[width=80mm]{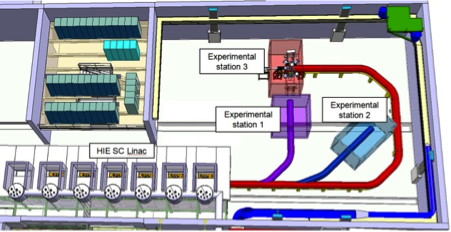}
   \caption{Layout of the HEBT lines in the ISOLDE experimental hall.}
   \label{HEBT_hall}
\end{figure}

The HEBT design~\cite{ipac12_op,ipac12_bd} is highly standardised with only a single type of quadrupole and dipole magnet needing procurement; the exceptions are the 22.5~deg dipoles on the transfer line to the second experimental station that are already procured and being refurbished. Each 90~deg bend is a standardised double-bend achromat consisting of two 45~deg dipoles and a single dispersion suppressing quadrupole. The HEBT is based on a periodic system of doublet cells with a unit length equal to the length of the high-$\beta$ cryomodule. As a result, the installation of the staged upgrade can occur in a modular fashion by the replacement of doublet cells with cryomodules. The HEBT will be extended in a second stage of the installation foreseen during 2017/18 to accommodate a third experimental station.  Some space has been reserved behind the third experimental station for the installation of a magnetic spectrometer in the future. The layout of the HEBT is shown in Figure~\ref{HEBT_hall}.
Vacuum in the low $10^{-7}~\textrm{mbar}$ range is obtained by turbomolecular pumps, backed by dry, hermetical multiroots, connected to the common effluent collector and the air ventilation evacuation. The turbopumps are installed at dedicated ports on the diagnostic boxes, which constitute the major source of length-specific outgassing. Vacuum sectorisation is kept minimal, with the goal of separating in distinct vacuum sectors the different beam lines, as well as to cut the experimental areas from the beam transfer sections.

\begin{figure}[ht]
   \centering
   \includegraphics[width=80mm]{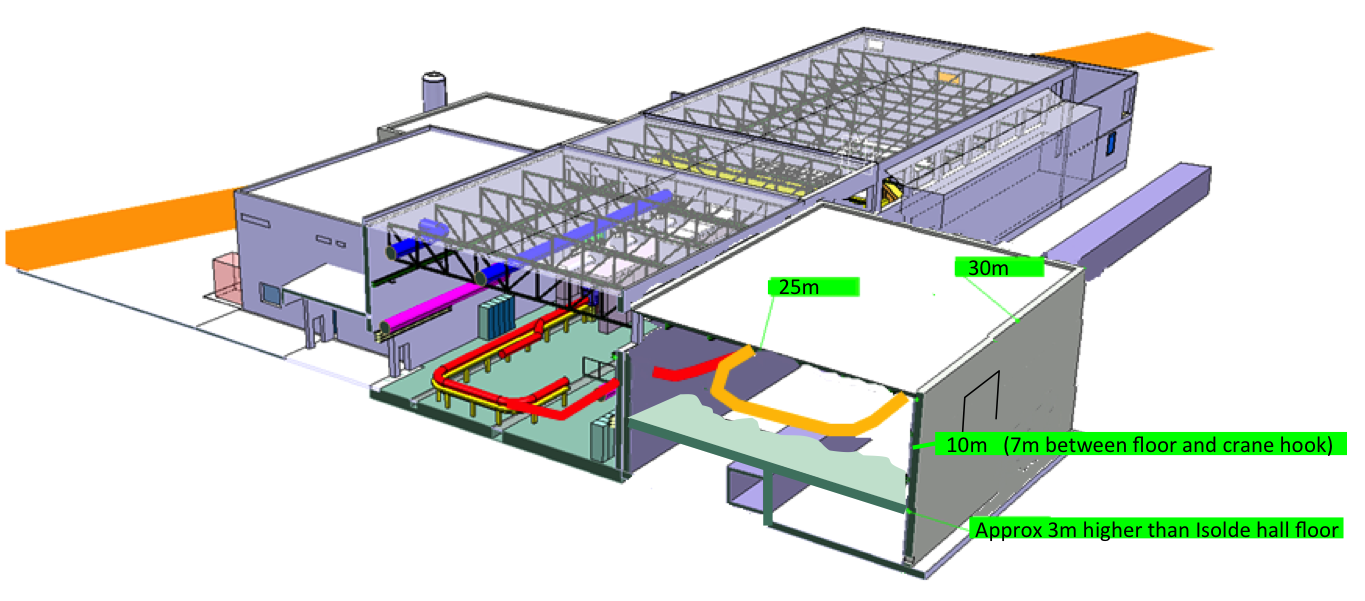}
   \caption{First-floor extension for the TSR behind the experimental hall (shown in orange) connecting to the HEBT via the inclined beam line (shown in red).}
   \label{tsr}
\end{figure}

The extension of the HEBT in the final stage permits a connection to the TSR storage ring that is expected to become operational at CERN during this period, see Figure~\ref{tsr}.

\section{SAFETY ASPECTS}

The main hazards that will be introduced by the linac upgrade come from X-rays, neutrons and the cryogenic system. A radiation measurement station has been integrated into the single-cavity test-stand to collect dose rate data specific to the HIE-ISOLDE cavities. These data will be used to confirm the sizing of the tunnel shielding. An access system with key is already used to prevent access during linac operation. When one key is removed from a key box the RF is stopped and the amplifiers cannot be restarted until all keys are back in a key box~\cite{voulot_safety}. A similar system is foreseen for the HIE-ISOLDE tunnel access. 

An incident beam of heavy ions at an energy of 10~MeV/$u$ is higher than the Coulomb barrier allowing neutron production. A dose rate of 0.1~$\textrm{mSv h}^{-1}~\textrm{pnA}^{-1}$  is expected at 1~m, at an angle of 90~deg from vicinity of the beam loss~\cite{health_physics}. During operation and setting-up, a maximum beam intensity of $\sim$ppA is expected in the experimental hall and free access will be possible. 

Redundant safety valves will be installed on the He tanks of the cryomodules to mitigate for the cryogenic hazard associated with the 900~litres of LHe inside the six cryomodules. These safety valves will be routed outside the tunnel in order to reduce exposure to an oxygen deficiency hazard during technical access to the tunnel while the cryomodules are cold.

Finally, pumping on radioactive beams implies a risk of radioactive contamination of the vacuum system, demanding an entirely hermetical system, with collection and evacuation of the effluent gases before controlled release to the atmosphere.

\section{PLANNING AND SCHEDULING}

If the carefully planned schedule is respected, the various experiments currently receiving beams from ISOLDE will not suffer from the upgrade works. Civil engineering work and installation of the main services such as power, ventilation and cooling will take place while the ISOLDE facility is running for the experiments. To guarantee a minimum of perturbation to the operation of the facility, the main services will be connected and the existing services modified only during the CERN Long Shutdown. Civil engineering work inside the experimental hall, such as the construction of the new tunnel, as well as the move of the existing Miniball experiment to its new position will also be carried out during this period of shutdown.

Start-up of the low energy (60~keV) part of the ISOLDE facility, excluding the REX post-accelerator, is foreseen for April 2014 as normal. However, at that moment the HIE-ISOLDE linac and HEBT line will still be under construction with the installation of the first two high-$\beta$ cryomodules and the transfer line elements running into the summer of 2014. Beam commissioning at 5.5~MeV/$u$ is planned for the end of the 2014. The remaining two high-$\beta$ cryomodules will be installed in a second stage in 2016, increasing the beam energy to 10~MeV/$u$, together with an additional bend in the HEBT providing the users with a third experimental station (and making possible the connection of the TSR storage ring).  A pre-bunching scheme before the RFQ with chopper and the two low-$\beta$ cryomodules completing the HIE-ISOLDE linac are foreseen for a later stage after 2017.

\section{OUTLOOK}

The main technological options of the linac are now fixed and most components are in their final design or prototyping phase. The HEBT design has been finalised and the components are now in the specification or prototyping phase. The infrastructure installation is under way and should be completed by mid-2013. The cryogenic lines and first phase of the linac and transfer lines installation is planned for 2014. The linac commissioning at 5.5~MeV/$u$ is foreseen for late 2014 and the first physics should take place in early 2015.

\section{ACKNOWLEDGMENTS}

This paper summarises the work of several teams: The ISOLDE Collaboration, the HIE-ISOLDE project team, and numerous groups at CERN within the accelerator and technology sector.

We acknowledge funding from the Swedish Knut and Alice Wallenberg Foundation (KAW 2005-0121) and from the Belgian Big Science program of the FWO (Research Foundation Flanders) and the Research Council K.U. Leuven.

We would also like to acknowledge the receipt of fellowships from the CATHI Marie Curie Initial Training Network (EU-FP7-PEOPLE-2010-ITN Project number 264330) and the COFUND-CERN Marie Curie Fellowship programme.

Support from the Spanish Programme ÒIndustry for ScienceÓ from CDTI is also acknowledged.

\end{document}